\documentclass[apj]{emulateapj}

\newcommand{\etal}{{et al.~}}
\newcommand{\msunh}{\>h^{-1}\rm M_\odot}

\newcommand{\mpch}{\>h^{-1}{\rm {Mpc}}}

\newcommand{\calC}{{\cal C}}
\newcommand{\rmag}{\>^{0.1}{\rm M}_r-5\log h}

\newcommand{\rp}{r_{\rm p}}
\def\gtsima{$\; \buildrel > \over \sim \;$}
\def\ltsima{$\; \buildrel < \over \sim \;$}
\def\gta{\lower.7ex\hbox{\gtsima}}
\def\lta{\lower.7ex\hbox{\ltsima}}

\shorttitle{The red dwarf galaxies}
\shortauthors{Wang et al.}

\begin{document}

%%%%%%%%%%%%%%%%%%%%%%%%%%%%%%%%%%%%%%%%%%%%%%%%%%%%%%%%%%%%%%%%%%%%%%%%%%

\title{The nature of red dwarf galaxies}

\author{Yu Wang\altaffilmark{1},
  Xiaohu Yang\altaffilmark{1},
  H.J. Mo\altaffilmark{2},
  Frank C. van den Bosch\altaffilmark{3},
  Neal Katz\altaffilmark{2},
  Anna Pasquali\altaffilmark{3},
  Daniel H. McIntosh\altaffilmark{4},
  Simone M. Weinmann\altaffilmark{5}}

\altaffiltext{1}{Key Laboratory for Research in Galaxies and Cosmology,
  Shanghai Astronomical Observatory; the Partner Group of MPA; Nandan Road 80,
  Shanghai 200030, China; E-mail: yuwang@shao.ac.cn}

\altaffiltext{2}{Department of Astronomy, University of Massachusetts, Amherst
  MA 01003-9305}

\altaffiltext{3} {Max-Planck-Institute for Astronomy, K\"onigstuhl 17, D-69117
  Heidelberg, Germany }

\altaffiltext{4}{Department of Physics, 5110 Rockhill Road, University of
  Missouri-Kansas City, Kansas City, MO 64110, USA}

\altaffiltext{5}{Max-Planck-Institut f\"ur Astrophysik,
  Karl-Schwarzschild-Strasse 1, 85748 Garching, Germany}

%%%%%%%%%%%%%%%%%%%%%%%%%%%%%%%%%%%%%%%%%%%%%%%%%%%%%%%%%%%%%%%%%%%%%%%%%%

\begin{abstract} Using dark matter halos traced by galaxy groups selected from
the Sloan  Digital Sky Survey Data  Release 4, we  find that about 1/4  of the
faint  galaxies  ($\rmag >-17.05$,  hereafter  dwarfs)  that  are the  central
galaxies  in their own  halo are  not blue  and star  forming, as  expected in
standard models of  galaxy formation, but are red.  In contrast, this fraction
is  about 1/2  for  dwarf satellite  galaxies.   Many red  dwarf galaxies  are
physically associated with more massive halos.  In total, about $\sim 45$\% of
red dwarf galaxies reside in  massive halos as satellites, while another $\sim
25$\% have a spatial distribution that is much more concentrated towards their
nearest massive  haloes than  other dwarf galaxies.   We use mock  catalogs to
show  that  the  reddest   population  of  non-satellite  dwarf  galaxies  are
distributed within  about 3  times the virial  radii of their  nearest massive
halos.   We suggest  that  this population  of  dwarf galaxies  are hosted  by
low-mass halos that have passed  through their massive neighbors, and that the
same environmental  effects that  cause satellite galaxies  to become  red are
also responsible for the red colors of this population of galaxies.  We do not
find any  significant radial  dependence of the  population of  dwarf galaxies
with the  highest concentrations, suggesting that the  mechanisms operating on
these galaxies affect color more  than structure.  However, over 30\% of dwarf
galaxies are red and isolated and their origin remains unknown.
\end{abstract}

%%%%%%%%%%%%%%%%%%%%%%%%%%%%%%%%%%%%%%%%%%%%%%%%%%%%%%%%%%%%%%%%%%%%%%%%%%

\keywords{dark matter  - large-scale structure of the universe - galaxies:
halos}

%%%%%%%%%%%%%%%%%%%%%%%%%%%%%%%%%%%%%%%%%%%%%%%%%%%%%%%%%%%%%%%%%%%%%%%%%%

\section{Introduction}
\label{sec:intro}

In the current paradigm of galaxy formation, it is believed that virtually all
galaxies initially form  as `disks' owing to the cooling  of gas with non-zero
angular momentum in  virialized dark matter haloes. This  smooth gas accretion
dominates  the galactic  gas supply  and hence  the fuel  for  star formation.
Galaxies that  reside in the  centers of lower  mass halos, those  with masses
less than  $M_{halo} \lta  3\times 10^{11} \msunh$,  accrete gas  through very
efficient  cold  mode accretion,  i.e.  gas that  is  never  heated (Keres  et
al. 2005,  Keres et  al. 2008).   The central galaxies  that reside  in larger
halos accrete their  gas through the classic, but less  efficient, hot mode of
accretion where  the gas is shock  heated to near the  virial temperature near
the virial  radius and then  must cool to  be accreted by the  central galaxy.
Hence the  naive expectation would be  that dwarf galaxies  should be actively
star forming and blue.

However, when a small halo is accreted by a larger halo, i.e.  when it becomes
a  subhalo,  the central  galaxy  that  formed in  the  small  halo becomes  a
satellite galaxy and may experience a number of environmental effects that may
change its  properties.  For instance,  the diffuse gas  originally associated
with  the  subhalo  may  be  stripped,  thus  removing  the  fuel  for  future
star-formation  (e.g.   Larson,  Tinsley  \& Caldwell  1980).   This  process,
referred to as strangulation (Balogh \&  Morris 2000), can result in a gradual
decline of the  star formation rate in the satellite  galaxy, making it redder
with  the passage of  time.  If  the external  pressure is  sufficiently high,
ram-pressure  stripping  may  also be  able  to  remove  the entire  cold  gas
reservoir of the satellite (e.g.  Gunn \& Gott 1972), causing a fast quenching
of its  star formation.  A satellite  galaxy is also subject  to tidal heating
and stripping and  galaxy harassment (Moore et al 1996),  which may also cause
the  satellite to  lose  its fuel  for  star formation.   These processes  are
believed  to have  played  an important  role  in the  evolution of  satellite
galaxies, and to  be responsible, to a large extent,  for the relation between
galaxy  properties  and  their  environment. Indeed,  satellite  galaxies  are
generally  found to  be redder  and  somewhat more  concentrated than  central
galaxies  with similar  stellar masses  (e.g.  van  den Bosch  et  al.  2008a;
Weinmann et al.  2008; Yang et al.  2008b; c; Guo et al. 2009).

Furthermore, recent investigations  based on cosmological $N$-body simulations
have shown  that a significant fraction  of dark matter haloes  that are close
to, but  beyond the  virial radius  of, a more  massive neighboring  halo, are
physically connected  to their neighbor.  As  shown by Lin et  al.  (2003) and
more  recently  by  Ludlow  et  al.   (2008), some  low-mass  halos  are  {\it
physically} associated  with more massive halos,  in the sense  that they were
once subhalos  within the virial radii  of these more  massive progenitors and
have subsequently been ejected.  This  population of halos was found to extend
beyond three times the virial radii  of their host halos, and represents about
$10\%$ of the entire population of low-mass halos (Wang, Mo \& Jing 2008).  If
galaxies have managed to form in the progenitors of these ejected halos, it is
likely that  the same environmental processes operating  on satellite galaxies
may also  have affected the properties  of these galaxies.   In particular, we
would  expect the  presence of  a red  population of  faint galaxies  that are
closely associated with massive halos that once hosted them.

Galaxies are observed to be bimodal in the color-magnitude plane: red galaxies
with  very little  star formation  (the red  sequence) and  blue  star forming
galaxies that are typically disky (the blue cloud) (e.g. Kauffman et al. 2003,
Baldry  et  al.  2004)  Extrapolating  the observed  division  line  (Yang  et
al.  2008a) to  dwarf galaxies  we surprisingly  find that  for  central dwarf
galaxies in the SDSS, with $r$-band magnitudes between -14.46 and -17.05, just
over 1/4 are red.   In this paper we will investigate the  nature of these red
dwarf galaxies.   Quantifying the spatial  distribution of this  population of
galaxies is  clearly important, because it  allows us to  determine whether or
not they  can be  explained as  a population of  satellite galaxies  that were
ejected from larger halos.

In this  paper, we use the galaxy  group catalogue constructed by  Yang et al.
(2007)  from  the  Sloan  Digital   Sky  Survey  Data  Release  4  (SDSS  DR4;
Adelman-McCarthy  \etal  2006) to  study  the  distribution  of central  dwarf
galaxies around massive halos.  The structure of this paper is as follows.  In
\S\ref{sec_data} we briefly describe the  criteria used to select galaxies and
galaxy  groups.  In \S\ref{sec_analyze}  we study  the radial  distribution of
dwarf  galaxies around  their nearest  neighbor  halos and  its dependence  on
galaxy color  and concentration. Some  systematic effects that may  change our
results are discussed in \S\ref{sec_systematics}.  In \S\ref{sec_mock}, we use
mock catalogues to  test the reliability of our results  and to quantify their
implications.   Finally, in  \S\ref{sec_discussion}, we  present  some further
discussion regarding our results.

\section{Observational Data}
\label{sec_data}

\subsection{Samples of Galaxy Groups}

Our analysis uses the galaxy group catalogues of Yang \etal (2007), which were
constructed from the New York University Value-Added Galaxy Catalog (NYU-VAGC,
see Blanton \etal 2005b) based on  the Sloan Digital Sky Survey Data Release 4
(SDSS DR4;  Adelman-McCarthy \etal  2006).  Only galaxies  in the  Main Galaxy
Sample with redshifts in the range $0.01 \leq z \leq 0.20$ and with a redshift
completeness $\calC  > 0.7$  were used.  Three  sets of group  catalogues were
constructed using a  modified version of an adaptive  halo-based group finder,
which was optimized  to assign galaxies into groups  according to their common
dark matter halos (Yang \etal 2005).   For our study here, we use group sample
II, in  which only galaxies  with spectroscopic redshifts (either  provided by
the SDSS or taken from alternative  surveys) are used. We have tested, though,
that  using group  sample III,  which also  includes galaxies  that  have been
missed owing to fiber collisions, does not have a significant impact on any of
our results.

For each group in the catalogue, Yang \etal (2007) estimated the corresponding
halo mass using either the ranking of its characteristic luminosity (this mass
is denoted by  $M_L$) or using the  ranking of its stellar mass  (this mass is
denoted by  $M_S$).  Throughout this paper,  we use $M_S$ as  our halo masses.
We  have also  tested that  using $M_L$  instead does  not change  any  of our
results.  As described in Yang \etal (2007), the characteristic luminosity and
stellar  mass of a  group are  defined to  be the  total luminosity  and total
stellar  mass of  all group  members, respectively,  with $\rmag  \leq -19.5$.
Thus, groups whose member galaxies are all fainter than $\rmag = -19.5$ cannot
be assigned halo masses according to  the ranking.  For these groups, the halo
masses are estimated in the following  way.  In Yang \etal (2008b) it is shown
that the  stellar masses of central  galaxies are tightly  correlated with the
masses of their host haloes.  The mean of this relation is well described by
\begin{equation}\label{eq:Ms_fit}
M_{\ast} = M_0
\frac { (M_h/M_1)^{\alpha +\beta} }{(1+M_h/M_1)^\beta } \,,
\end{equation}
where $M_{\ast}$  and $M_h$ are the  central galaxy stellar mass  and the host
halo mass of  the group, respectively, and ($\log  M_0$, $\log M_1$, $\alpha$,
$\beta$) = (10.306, 11.040, 0.315, 4.543).  For groups that cannot be assigned
a halo  mass according  to the stellar-mass  (luminosity) ranking, we  use the
above relation  to obtain  $M_h$ through the  stellar masses of  their central
galaxies.

\subsection{Galaxy Samples}
\label{sec:samples}
\begin{deluxetable*}{lccccccccc}
\tabletypesize{\scriptsize}
%\rotate
\tablecaption{Galaxy Samples \label{tab1}} \tablewidth{0pt} \tablehead{ID &
  $\rmag$ & $N_{\rm total}$ & $N_{\rm cent}$ & $N_{\rm sat}$ &
  $f_{\rm red,cent}$ &  $f_{\rm red,sat}$ & $f^b_{\rm red,cent}$ & $f^b_{\rm
    red,sat}$ &  $f^b(\rp/R_{180})\le 3$ \\
  (1) & (2) & (3) & (4) & (5) & (6) & (7) & (8) & (9) & (10)} \startdata
S1       & (-14.46,-16.36] & 1500 & 1103 & 397 & 13.69\% & 37.53\% & 33.79\% &
         53.42\% & 34.60\%\\
S2       & (-16.36,-16.78] & 1500 & 1081 & 419 & 13.69\% & 36.28\% & 22.50\% &
         47.02\% & 47.46\%\\
S3       & (-16.78,-17.05] & 1500 & 1008 & 492 & 10.20\% & 40.24\% & 19.51\% &
         52.64\% & 47.72\%\\
S1+S2+S3 & (-14.46,-17.05] & 4500 & 3192 & 1308 & 12.59\% & 38.15\% & 25.49\% &
         51.08\% & 41.63\% \\
\cline{1-10}\\
S4 & & 1500 & 1080 & 420 & 13.51\% & 37.02\% \enddata

\tablecomments{Column 1 indicates  the sample ID. Column 2  lists the absolute
  magnitude range  of each  sample.  Columns  3 to 5,  indicate the  number of
  total, central and satellite galaxies in each sample, respectively.  Columns
  6  and  7 list  the  red fractions  among  central  and satellite  galaxies,
  respectively, where the  red galaxies are defined to be  the reddest 20\% of
  all the galaxies. Column 8 lists  the red fraction of central dwarf galaxies
  where the red galaxies are defined by extrapolating the division between red
  sequence and blue cloud galaxies from  Yang et al.  (2008a).  Column 9 lists
  also  this red fraction,  but for  the satellite  dwarf galaxies.  Column 10
  lists the  fraction of those (in  Column 8) central red  dwarf galaxies that
  have $\rp/R_{180}\le 3$. }
\end{deluxetable*}

Group catalogue II consists of 369447 galaxies, which are assigned into 301237
groups.  The  majority, 271420, of the  groups contain only  one member, i.e.,
all  of them  are the  central galaxies  of the  groups.  The  remaining 98027
galaxies are in groups  with more than one member, and 29817  of them are {\it
central} galaxies  (the brightest one in  each group).  We refer  to the other
68210 galaxies as {\it satellites} (Yang et al.  2007).

From  our galaxy  sample,  we select  three  subsamples of  dwarf galaxies  as
follows.  We  rank order all  galaxies according to their  absolute magnitudes
(in the $r$-band, $K$- and evolution- corrected to redshift $z=0.1$), starting
with the  faintest galaxy. The 1500  galaxies with the highest  rank (i.e. the
1500 faintest galaxies) make up our first sample, called S1, The galaxies with
ranks  1501-3000 make  up sample  S2, and  those with  ranks  3001-4500 sample
S3. Table~\ref{tab1}  lists the (sequential) absolute magnitude  ranges of all
three samples, as well as the  numbers of central and satellite galaxies. Note
that all the galaxies in S1, S2  and S3 are fainter than $\rmag = -17.05$.  In
what  follows  we refer  to  all  galaxies in  these  three  samples as  dwarf
galaxies.

\begin{figure} \plotone{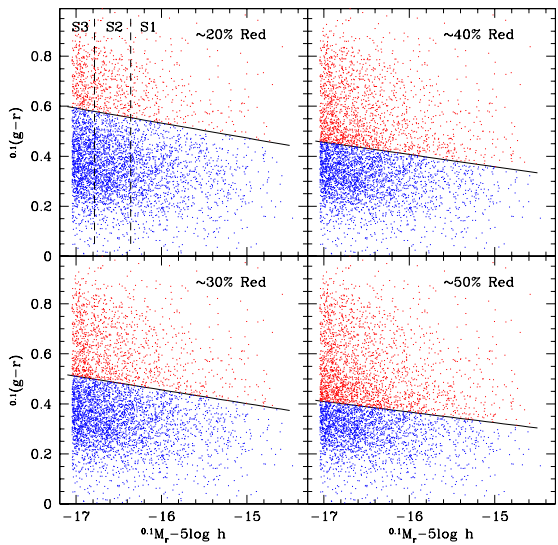}
  \caption{The  color-magnitude  distribution  of  dwarf  galaxies  (including
    central and  satellite galaxies).  In  four panels the dwarf  galaxies are
    separated  into  20\%,  30\%,   40\%,  50\%  red  galaxies  as  indicated,
    respectively.  The vertical dashed lines  in the upper-left panel show the
    separation criteria of samples S1, S2 and S3.} \label{fig:col}
\end{figure}

To  study how the  spatial distribution  of dwarf  galaxies depends  on galaxy
color, we  separate each of  the samples,  S1, S2, and  S3, into red  and blue
subsamples. In particular, we define a color cut
\begin{equation}\label{colcut}
^{0.1}(g-r) = a + b~(\rmag)\,,
\end{equation}
and we  adjust the parameters  $a$ and  $b$ such that  samples S1, S2,  and S3
roughly have the  same fractions of galaxies, $f_{\rm  red}$, redder than this
particular cut. We  consider four values for $f_{\rm  red}$: 20\%, 30\%, 40\%,
and  50\%,  for  which  we  obtain  [a,  b]=[-0.421,-0.060],  [-0.423,-0.055],
[-0.375,-0.049] and  [-0.313,-0.043], respectively.   Thus, if $f_{\rm  red} =
20$\% it means  that the red subsamples of  S1, S2 and S3 each  consist of the
20\%  reddest galaxies  in their  particular samples,  etc.  Fig~\ref{fig:col}
shows the color-magnitude  relations of galaxies in S1,  S2 and S3 (delineated
by vertical dashed  lines).  The four panels correspond  to the four different
values  of $f_{\rm  red}$, as  indicated,  and the  solid line  in each  panel
corresponds to the color cut of Eq.~(\ref{colcut}) used in each case.

In Table~\ref{tab1}  we list,  for each sample,  the red fractions  of central
($f_{\rm  red,cent}$) and  satellite galaxies  ($f_{\rm red,sat}$).   Here red
galaxies are defined to be the reddest 20\% of all galaxies (both centrals and
satellites) in our sample of dwarf galaxies.  Clearly, dwarf galaxies that are
satellites have a  much higher red fraction than  central dwarf galaxies.  For
comparison, extrapolating the observed  division line between the red sequence
and the blue cloud  from Yang et al. (2008a) down to  dwarf galaxies one would
find that  33.8\%, 22.5\%, and 19.5\% of  the central galaxies in  the S1, S2,
and S3  samples, respectively, were red.   While in total, there  are more red
central  dwarfs than  red satellite  dwarfs, which  is so  far  not predicted,
e.g.  by halo  occupation  models (Brown  et  al. 2008).  Note that  different
definition of  red galaxies may change  these fractions (e.g.  with respect to
the galaxies  of similar stellar masses),  however not the  general results we
find in this paper.

\section {The distribution of central dwarfs}
\label{sec_analyze}

\begin{figure} \plotone{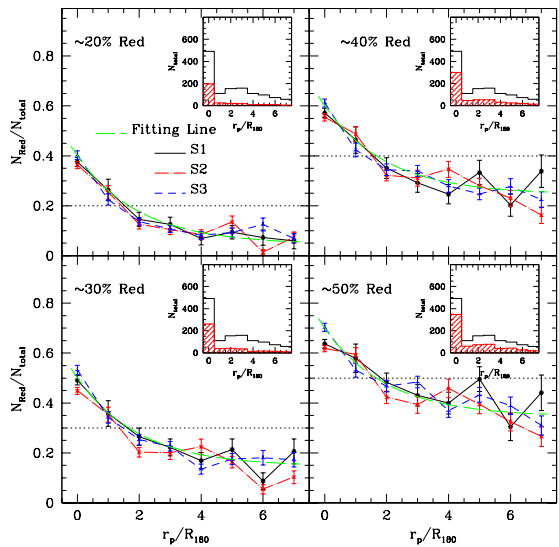}
  \caption{Fractions of the red central  galaxies near more massive halos as a
    function of  projected distance $\rp/R_{180}$.   The four panels  show the
    results  of  the color  subsamples  with 20\%,  30\%,  40\%  and 50\%  red
    galaxies, respectively.   The related color  subsample separation criteria
    are shown in  Fig.  \ref{fig:col}.  The different lines  correspond to the
    different  samples  as indicated.   The  fit  long-dashed,  green line  is
    described in  the text. For comparison,  we show the fraction  of red {\it
      satellite} galaxies  that are within  the same luminosity ranges  as the
    central galaxies  at $\rp/R_{180}=0$. The  small window within  each panel
    shows the number counts of dwarf  central (or satellite) galaxies in S1 as
    a function of $\rp/R_{180}$: black  histogram for all galaxies and shaded,
    red histogram for red galaxies. } \label{fig:f_r}
\end{figure}

In this  section, we  investigate how central  dwarf galaxies  are distributed
with respect to their nearest more massive halo (i.e.  more massive than their
own halo).   Since the  distances of galaxies  based on redshifts  suffer from
redshift distortions, we separate the  distance between a central dwarf galaxy
and its  nearest more massive  halo into two  components: $\pi$, which  is the
separation along the line-of-sight, and  $\rp$, which is the separation in the
perpendicular direction.

For each  group in  the catalogue, we  use the  assigned halo mass,  $M_S$, to
estimate its halo radius, $R_{180}=[3M_{S}/(4\pi*180\bar{\rho})]^{1/3}$, which
follows from defining  the mean mass density within a halo  as $180$ times the
average density of the universe,  $\bar{\rho}$.  We search around each central
dwarf   galaxy,  within   a  line-of-sight   separation  $|\pi|   =  15\mpch$,
\footnote{Tests have shown that  changing the line-of-sight separation for the
search  from $|\pi|  \le 15\mpch$  to  $|\pi| \le  10\mpch$ or  to $|\pi|  \le
20\mpch$ does not  have a significant impact on any of  our results.}  for the
group  that has  (i) a  halo mass  larger  than that  of the  dwarf galaxy  in
question and (ii)  the lowest value of $\rp/R_{180}$  (with $R_{180}$ the halo
radius of  the group).   The central  galaxy is then  said to  be at  a scaled
`distance' $\rp/R_{180}$ from  a massive halo, and the halo  is referred to as
the nearest halo of the galaxy.  We use this scaled distance because $R_{180}$
is the  only important length  scale related to  the dynamics of  a virialized
halo.

\subsection{Color Dependence}

Fig.~\ref{fig:f_r} shows  the fraction $N_{\rm red}/N_{\rm  total}$ of central
dwarf  galaxies  that   are  red  as  a  function   of  the  scaled  distance,
$\rp/R_{180}$,  to their  nearest halos.  Here  $N_{\rm total}$  is the  total
number  of dwarf  galaxies  in each  sample  (S1, S2  or S3)  in  that bin  of
$\rp/R_{180}$, and $N_{\rm red}$ is the  number of those galaxies that are red
according to the criterion used. The  numbers are shown in the small window in
each  panel (the black  histogram for  $N_{\rm total}$,  and the  red, hatched
histogram for  $N_{\rm red}$).  The four  panels show the  results for $f_{\rm
red} =  20$\%, 30\%, 40\% and 50\%.   In each panel, the  different lines show
the results  obtained for the three samples,  S1, S2, and S3,  as indicated in
the  upper-left  panel.   The  error-bars  are obtained  using  100  bootstrap
resamplings  (Barrow, Bhavsar,  \& Sonoda  1984; Mo,  Jing \&  B\"orner 1992).
Galaxies are counted in bins  specified by $N-0.5 \leq \rp/R_{180} \leq N+0.5$
(for  $N=2, 3,... ,7$),  and $0\leq  \rp/R_{180} \leq  N+0.5$ for  $N=1$.  For
comparison, the data point at  $\rp/R_{180} =0$, indicates the fraction of red
{\it  satellite} galaxies  within the  same  luminosity range  as the  central
galaxies.   Note that  the results  obtained  for S1,  S2, and  S3 are  almost
identical, indicating  that the spatial distribution of  dwarf galaxies around
massive haloes does not depend on their luminosities.  However, if we consider
much  brighter galaxies,  e.g. at  $\rmag \sim  -18.5$, the  radial dependence
starts to level off.

There  is a  clear  trend that  the  fraction of  red  central dwarf  galaxies
increases with decreasing  scaled distance to the nearest  halo.  The fraction
of the  20\% reddest population at  $\rp/R_{180}\ga 4$ is around  5\% to 10\%,
increases systematically to  $\sim 25$\% at $\rp/R_{180}\sim 1$,  and to $\sim
40$\% at  $\rp/R_{180} = 0$ for  the satellite galaxies.  For  the other three
cases (with $f_{\rm  red} =30$\%, 40\% and 50\%),  the fraction also decreases
with $\rp/R_{180}$, but  reaches a higher level at  large $\rp/R_{180}$.  This
indicates  that the less  red galaxies  in these  subsamples are  not strongly
associated  with massive  halos.  We  quantify  the association  of red  dwarf
galaxies with massive halos by fitting  the data obtained from S3 shown in the
four  panels  simultaneously with  a  function $f=a+b\times\exp(-x/2)$,  where
$x=\rp/R_{180}$.   In the  fit we  subtract a  constant of,  $0.1$,  $0.2$ and
$0.3$,  from  the  data  for  the  30\%, 40\%  and  50\%  reddest  subsamples,
respectively, to account for the component that is not closely associated with
massive halos.  The best fit  results, with $a=0.045$ and $b=0.356$, are shown
in each panel of Fig.  \ref{fig:f_r} as the green, long-dashed lines.  We have
also checked the radial distribution  of the central dwarf galaxies when using
$f_{\rm red} = 15$\%  to define the subsample of red dwarfs.  In this case, we
find less  than a  5\% decrease  in the red  fraction at  large $\rp/R_{180}$,
indicating that the  5\% least red galaxies in the  20\% reddest subsample are
not randomly but more closely  distributed relative to the massive halos. Thus
the overall  results suggest that the  $15\%$ - $20\%$  reddest dwarf galaxies
are  quite  distinct from  the  other  dwarfs, in  that  they  reveal a  clear
preference to reside close to their  nearest more massive dark matter halo. In
addition,  the non-zero  red fraction  at large  $\rp/R_{180}$  indicates that
there is a  $\sim 5\%$ tail of red dwarfs  randomly distributed throughout the
background population,  especially in  the voids, due  to some  processes that
shut down the star formation. A similar trend has also been found by Cooper et
al. (2007)  from the DEEP2  survey, however for  more massive galaxies  in low
density regions at  higher redshifts. In \S\ref{sec_mock}, we  use mock galaxy
redshift  surveys  constructed   from  cosmological  $N$-body  simulations  to
quantify this connection.

\subsection{Concentration Dependence}

Apart from a  color dependence, we also check  whether the radial distribution
of dwarf galaxies with respect to their nearest more massive dark matter halos
depends on their surface brightness profiles.  To this end, we split our dwarf
galaxies into  two subsamples  according to the  value of  their concentration
parameter $C=r_{90}/r_{50}$.   Here $r_{90}$ and  $r_{50}$ are the  radii that
contain 90  and 50 percent of  the Petrosian $r$-band  flux, respectively.  As
shown in Strateva \etal (2001), $C$ is a reasonable proxy for the Hubble type,
with $C>2.6$  corresponding to  early-type galaxies.  We,  therefore, separate
galaxies  into  high  ($C>2.6$)   and  low  ($C\le  2.6$)  concentrations,  as
illustrated in  the lower-right panel of  Fig.~\ref{fig:concen}.  Roughly 20\%
of the dwarf galaxies thus end up in the high-concentration subsample.

The lower-right  panel of Fig.~\ref{fig:f_con} shows the  fraction of galaxies
in this high-concentration  subsample as a function of  the scaled distance to
the  nearest  more  massive  halo.   Unlike the  reddest  galaxies,  the  most
concentrated galaxies  have a radial distribution  that is similar  to that of
the total  population of central  dwarf galaxies.  Note however,  for brighter
galaxies,  especially   in  and  around   clusters,  there  is  a   so  called
morphology-radius relation  (e.g., Dressler et  al. 1997), which  according to
Park \& Hwang  (2008) may be largely induced by the  interaction of the target
galaxy with its nearest (early-type) neighbor galaxy.

\section{Systematics}
\label{sec_systematics}

Before we proceed to study the origin of the central red dwarf galaxies, there
are a number of issues that need to be addressed. An obvious worry is that the
group  finder is  not perfect,  and has  misclassified a  number  of satellite
galaxies as central galaxies.  We will  discuss this issue in more detail with
the  help of  mock galaxy  and group  catalogs in  \S\ref{sec_mock}.   In this
section we discuss systematics that can be addressed without the need for mock
catalogs.

\subsection{Stellar Mass Dependence}
\begin{figure} \plotone{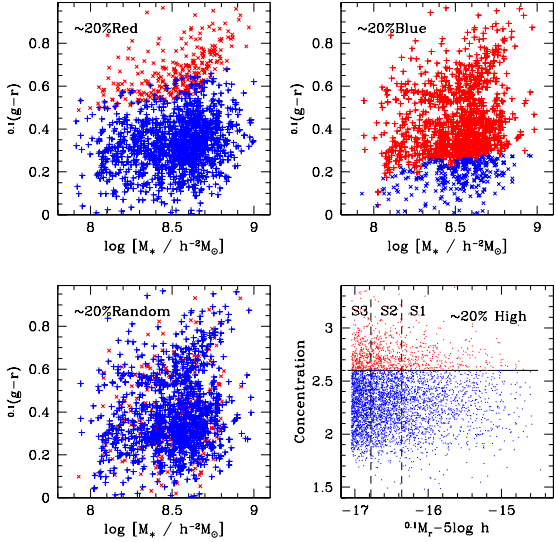}
  \caption{Upper-left panel:  the color-stellar mass distribution  of the 20\%
    red (`x'-crosses) and  80\% blue (`+'-crosses) dwarf galaxies  in terms of
    {\it similar  stellar masses}.  Upper-right panel: the  same set  of dwarf
    galaxies  as  in  the  upper-left  panel, but  separated  into  20\%  blue
    (`x'-crosses) and 80\% red  (`+'-crosses).  Lower-left panel: the same set
    of dwarf  galaxies as in the  upper-left panel, but  randomly selected and
    separated  into  20\% (`x'-crosses)  and  80\% (`+'-crosses)  populations.
    Lower-right panel:  the concentration-magnitude distribution  of the dwarf
    galaxies, which  are separated into $\sim  20\%$ high and  $\sim 80\%$ low
    concentration populations by $C=2.6$. } \label{fig:concen}
\end{figure}
\begin{figure} \plotone{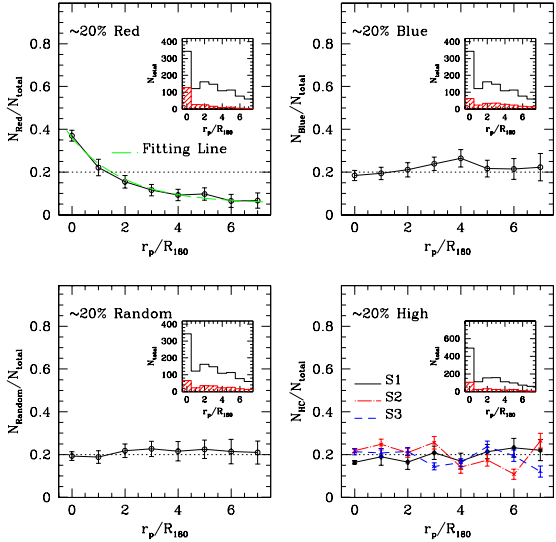}
  \caption{Similar  to   Fig.   \ref{fig:f_r},  but   with  related  subsample
    separation  criteria shown in  Fig.  \ref{fig:concen}.   Upper-left panel:
    fraction of the red central galaxies near more massive halos as a function
    of $\rp/R_{180}$, where the 20\% red population is defined with respect to
    the  similar stellar  mass galaxies.   Upper-right panel:  similar  to the
    upper-left  panel but  for the  20\% blue  population.   Lower-left panel:
    similar  to the  upper-left  panel  but for  the  20\% random  population.
    Lower-right  panel: fraction  of  the high  concentration $C>2.6$  central
    galaxies    near    more    massive     halos    as    a    function    of
    $\rp/R_{180}$. } \label{fig:f_con}
\end{figure}

For a given luminosity, redder galaxies  are expected to have a larger stellar
mass.  The color separation used above  may thus introduce a bias in the sense
that galaxies  in the  redder subsample are  systematically more  massive.  To
check  whether or  not  the color  distribution  we obtained  in the  previous
section is  robust when the dwarf  galaxies are selected in  a similar stellar
mass bin,  we construct  a controlled subsample  S4, where stellar  masses for
galaxies are  estimated using the  relation between the  stellar mass-to-light
ratio  and color  obtained by  Bell et  al.(2003). Note  that the  survey {\it
magnitude limit} of the SDSS  observation corresponds to a higher (lower) {\it
stellar  mass limit} for  the redder  (bluer) galaxies.   In general,  one can
construct a  stellar mass limit  sample (and hence  the subsample S4)  for all
(including both red and blue) galaxies  by adopting the stellar mass limit (as
a function  of redshift)  for the  reddest galaxies (see  Appendix of  van den
Bosch et  al. 2008b), which, however,  may significantly reduce  the number of
dwarf galaxies in our sample.  Instead, as a rough approximation, we construct
subsample S4 as  follows (with the hidden assumption that  if the galaxies are
complete  in  both  luminosity  and  stellar  mass  they  have  similar  color
distributions and thus similar red and blue fractions). First, we separate all
the dwarf galaxies  into red and blue populations: the  reddest 20\% being red
and the  rest being blue,  using the separation  line shown in  the upper-left
panel of  Fig ~\ref{fig:col}. Next,  for each of  the (1500$\times$20\%$=300$)
red  galaxies in  S1, we  randomly  select four  blue galaxies  from the  blue
population with stellar masses within  $\Delta \log M_{\ast}=0.025$ of the red
galaxy.  This yields a  blue control sample of (1500$\times$80\%$=1200$) dwarf
galaxies, which  has the  same stellar mass  distribution as the  20\% reddest
galaxies. The  control subsample  S4 so constructed  has exactly the  same red
fraction of dwarf  galaxies as S1, but now with respect  to blue galaxies with
similar stellar  masses.  The upper-left panel  of Fig.~\ref{fig:concen} shows
the color-stellar mass relation for S4, split into red and blue galaxies.

For comparison, we also form the following two subsamples from S4.  In one, we
randomly select 20\% of the galaxies from S4; in the other, we select the 20\%
bluest  galaxies that  have  the same  stellar  mass distribution  as all  the
galaxies  in  sample  S4.   The  color-stellar mass  relations  of  these  two
subsamples  are  shown  in  the  lower-left and  upper-right  panels  of  Fig.
\ref{fig:concen}, respectively.

The upper-left panel of Fig. \ref{fig:f_con} shows $N_{\rm red}/N_{\rm total}$
as a  function of  $\rp/R_{180}$ obtained using  sample S4.  Fitting  the data
again with  the function  $f=a+ b \times\exp(-x/2)$,  we obtain  $a=0.053$ and
$b=0.309$,  and the  corresponding model  is shown  as the  long-dashed curve.
This  dependence  on  the  scaled  distance, $x\equiv  \rp/R_{180}$,  is  only
slightly  weaker  than  that  for  the  corresponding  luminosity  sample  S1,
indicating that the bias caused by the stellar-mass difference between the red
and blue subsamples is not important.

The  upper-right  and lower-left  panels  of  Fig.   \ref{fig:f_con} show  the
results obtained for the 20\% bluest galaxies and for the 20\% random galaxies
(as  defined above).   For  these two  cases  there is  no significant  radial
dependence.  Although  one expects  such a lack  of radial dependence  for the
random subsample,  it does indicate  that there are no  significant systematic
errors in our  analysis.  The lack of a radial dependence  for the 20\% bluest
subsample is  due to the  fact that only  the $\sim 15-20\%$  reddest galaxies
reveal a radial distribution that is peaked towards smaller $\rp/R_{180}$.

\subsection{Dependence on the Mass of the Nearest Neighbor}

\begin{figure} \plotone{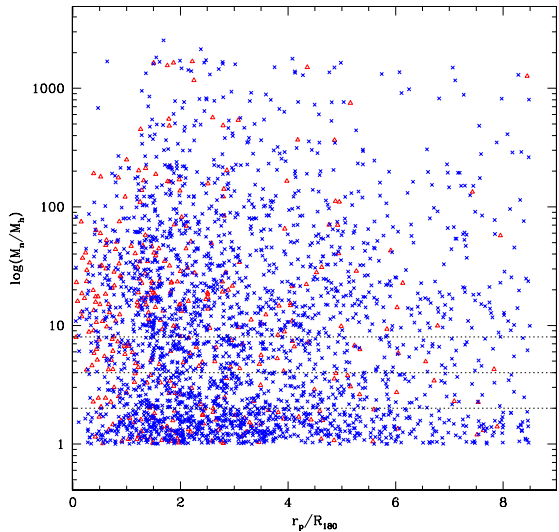}
  \caption{The nearest neighbor to host  halo mass ratio $M_n/M_h$ - projected
    distance  $\rp/R_{180}$  distribution of  the  dwarf  central galaxies  in
    samples S1+S2+S3. The triangles and  crosses show the red and blue central
    galaxies, respectively, where the red population is defined to be the 20\%
    reddest all galaxies.  }  \label{fig:mass_rp}
\end{figure}
\begin{figure} \plotone{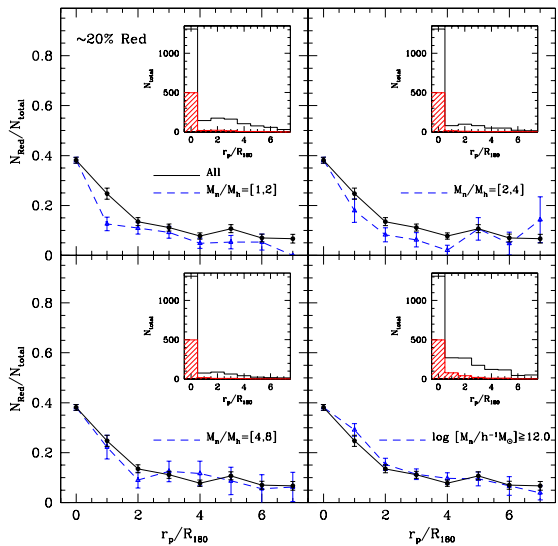}
  \caption{Similar  to Fig.  \ref{fig:f_r}, but  for all  galaxies  in samples
    S1+S2+S3 within different central-nearest halo systems.  In each panel the
    selection criteria,  $M_n/M_h$, is indicated.  Here the  results are shown
    for  the fraction  of the  20\% red  population. For  comparison,  in each
    panel, we  also show as  the symbols with  solid line the results  for all
    central-nearest halo systems.  } \label{fig:M_n}
\end{figure}

In our  analysis above, the ``nearest  more massive halo'' of  a central dwarf
galaxy  is defined  as the  halo with  a line-of-sight  separation  $|\pi| \le
15\mpch$ which  has (i) a  mass that  is more massive  than that of  the dwarf
galaxy,    and   (ii)    the    smallest   value    of   $\rp/R_{180}$    (see
section~\ref{sec_analyze}).   This implies  that  some of  these nearest  more
massive halos may  have masses that are only slightly larger  than that of the
dwarf galaxy under consideration.

In what follows  we use $M_h$ to refer  to the halo mass of  the central dwarf
galaxy, and $M_n$ to refer to the  mass of its nearest more massive halo.  For
our  combined sample  (S1 +  S2 +  S3)  the average  value of  $M_h$ is  about
$10^{10.9}\msunh$.   Fig.~\ref{fig:mass_rp}  shows the  ratio  $M_n/M_h$ as  a
function of the  scaled distance $\rp/R_{180}$ for all  central dwarf galaxies
in S1+S2+S3.  Here the results for  the 20\% reddest galaxies are shown as red
triangles, while the other 80\% are indicated by blue crosses. Note that there
is a very large amount of scatter  in $M_n/M_h$, ranging from unity to well in
excess of 1000.

It is  interesting to investigate whether  the color dependence  of the radial
distribution of central dwarfs with respect to their nearest more massive halo
depends on $M_n$. This can provide  valuable insight into the actual origin of
this color  dependence.  We  therefore proceed as  follows.  We  first combine
samples  S1+S2+S3,  and then  calculate  the  fraction  of central  red  dwarf
galaxies  that   belong  to   the  20\%  reddest   subsample  as  we   did  in
\S\ref{sec_analyze}. However,  now we only select systems  for which $M_n/M_h$
is  restricted  to  [1,2],  [2,4],  [4,8] or  $\log  [M_n/  \msunh]\ge  12.0$,
respectively.  The results are shown  in the four panels of Fig.~\ref{fig:M_n}
as indicated. For comparison we also show, in each panel, the results obtained
for all systems (i.e.  $M_n/M_h >  1$).  A comparison of all four panels shows
that  there is  a clear,  albeit  somewhat weak,  dependence of  the trend  on
$M_n/M_h$.  Overall,  the colors of  central dwarf galaxies are  most strongly
affected by nearest neighbor halos that are  more massive. In the case of $1 <
M_n/M_h \leq 2$ (upper-left panel),  the central dwarf galaxies have a $N_{\rm
red}/N_{\rm  total}$ that  is almost  independent of  $\rp/R_{180}$,  and much
lower  than that  of the  dwarf satellites.   On the  other hand,  the central
dwarfs that are distributed around  halos more massive than $10^{12.0} \msunh$
(lower-right panel), have  a radial dependence that is  somewhat stronger than
that for all systems.

\subsection{Survey Edge Effect}
\begin{figure} \plotone{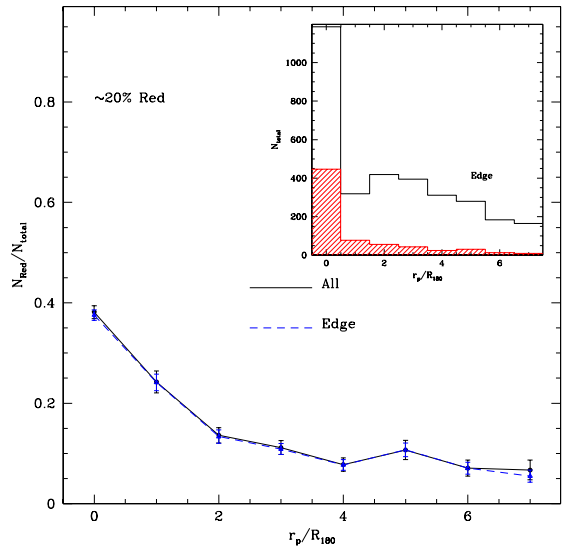}
  \caption{Similar to Fig. \ref{fig:f_r}, but  here we compare the results for
    all  galaxies  in samples  S1+S2+S3,  with  and  without edge  effects  by
    removing   the  central   dwarf  galaxies   that  are   near   the  survey
    edge. } \label{fig:edge}
\end{figure}

Since the  SDSS is  not a full-sky  survey, and  since our group  catalogue is
constructed using  only galaxies  with redshifts $0.01  \leq z \leq  0.2$, our
results may  be influenced by  edge effects of  the survey: for  central dwarf
galaxies near an edge of the  survey, there is an enhanced probability that it
is actually  a satellite (or central)  galaxy in a more  massive group (halo),
but for  which all other members  just happen to  lie beyond the edges  of the
survey. Although  we tried to take  these effects into  account when assigning
halo masses to our  groups (see Yang et al. 2007 for  details), it could still
be that  a significant fraction of  our central dwarf galaxies  are in reality
misclassified centrals or satellites owes to the survey geometry.

To check the  impact of these edge  effects, we follow Yang et  al.  (2007) by
measuring the edge parameter $f_{\rm edge}$.  For each central dwarf galaxy in
S1+S2+S3, we randomly  distribute $500$ points within a  radius $1\mpch$. Next
we apply  the SDSS DR4  survey mask and  remove those random points  that fall
outside of the region where the  completeness $\calC > 0.7$.  For each central
dwarf galaxy we then compute the number of remaining points, $N_{\rm remain}$,
and we  define $f_{\rm edge}=N_{\rm remain}/500$  as a measure  for the volume
around the  central dwarf galaxy that  lies within the survey  edges.  To test
the  impact of  edge effects  on our  measurements in  \S\ref{sec_analyze}, we
remove those  central dwarf galaxies  with $f_{\rm edge}\le 0.8$  (about 13\%)
and recalculate the radial distribution of the remaining central galaxies. The
result is shown in Fig.  \ref{fig:edge} (dashed line), compared to the results
for  all the  central dwarfs,  independent of  their value  of  $f_{\rm edge}$
(solid  line).    Clearly,  the  two  curves   are  almost  indistinguishable,
indicating that our results are not an artifact of survey edge effects.

\section{Test with mock samples}
\label{sec_mock}

\begin{figure} \plotone{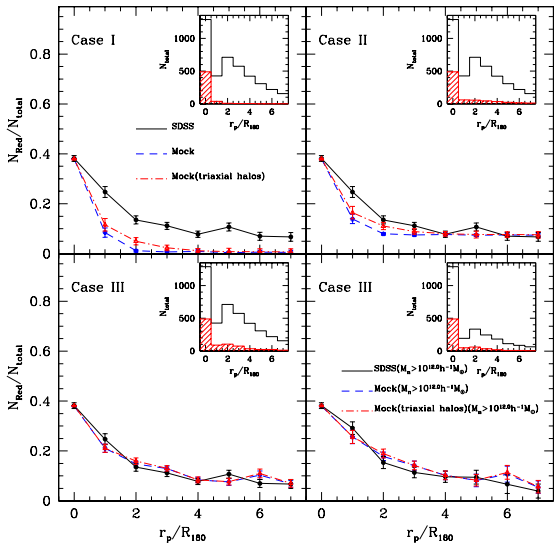}
  \caption{Similar to Fig \ref{fig:f_r}, but here we compare the observational
    and mock results.  The  upper-left, upper-right and lower-left panels show
    the  results  for all  host-nearest  halo  systems  using different  color
    models: Case  I, II and III  as indicated (see text).   In the lower-right
    panel, we show results for those central-nearest halo systems with $M_n\ge
    10^{12.0}\msunh$ using color model Case III but with different parameters.
    In  each  panel, the  symbols  connected  with  dashed lines  are  results
    obtained from  the mock  galaxy and group  catalogues where the  halos are
    assumed to be  spherical. The symbols connected with  dot-dashed lines are
    results obtained from the mock galaxy and group catalogues where the halos
    are  assumed  to  follow a  triaxial  Jing  \&  Suto (2002)  profile.  For
    reference, in  each panel we also  show, as the dots  connected with solid
    lines, the  results we obtained for  the SDSS samples  S1+S2+S3.  See text
    for details.  } \label{fig:mock}
\end{figure}

One  potential problem  with the  results presented  above is  that  the group
finder used  to identify  galaxy groups  is not perfect.   Hence, some  of the
dwarf  galaxies  classified as  central  galaxies  may  in fact  be  satellite
galaxies.   To test  the severity  of such  effects and  to quantify  the true
association between central dwarf galaxies  and their nearby massive halos, we
apply  the same  analysis to  mock samples  and compare  the results  with the
observational  data that  we have  obtained.  Here  we use  the mock  SDSS DR4
galaxy and  group catalogues that  are constructed by  Yang et al.   (2007) to
test the performance of the group  finder.  Following Yang et al.  (2004), the
mock  galaxy catalogue  is constructed  by  populating dark  matter haloes  in
numerical  simulations of  the standard  $\Lambda$CDM model  with  galaxies of
different luminosities, using the  conditional luminosity function (CLF) model
of  Cacciato et  al.  (2008).  The cosmological  parameters  adopted here  are
consistent with the  three-year data release of the  WMAP mission: $\Omega_m =
0.238$,  $\Omega_{\Lambda}=0.762$, $n_s=0.951$,  $h=0.73$  and $\sigma_8=0.75$
(Spergel et al.  2007).  This CLF describes the  halo occupation statistics of
SDSS galaxies, and accurately matches the SDSS luminosity function, as well as
the clustering and  galaxy-galaxy lensing data of SDSS  galaxies as a function
of their luminosity.  Next a mock redshift survey is constructed mimicking the
sky  coverage of  the SDSS  DR4  and taking  detailed account  of the  angular
variations in the magnitude limits and  completeness of the data (see Li \etal
2007  for details).  Finally  we construct  a group  catalogue from  this mock
redshift survey, using  the same halo-based group finder as  for the real SDSS
DR4.

To  test the  impact of  contamination on  our observational  results obtained
above, we consider three models for the distribution of red dwarf galaxies:
\begin{itemize}
\item Case I: Here we assume that a fraction, $f_{\rm red, sat}$, of true {\it
    satellite}  dwarf  galaxies are  red,  but  that  all true  central  dwarf
  galaxies are blue.
\item Case  II: Same as Case  I, but here  we assume that a  fraction, $f_{\rm
    red, cent}$, of true central dwarf galaxies are also red and have the same
  spatial distribution as blue central dwarfs.
\item  Case III: Similar  to Case  II, but  here we  assume that  $f_{\rm red,
    cent}$ depends on  the distance of the central galaxy  to its nearest more
  massive halo, according  to $f_{\rm red, cent}(r)= a+b\times\exp(-(y-1)/2)$.
  Here $a$ and $b$ are constants and $y=r/R_{180}$.
\end{itemize}
These models  are used to assign  a color to  each of the mock  dwarf galaxies
according to their  positions in real space and their  true membership of host
halos.

As for the  observational data, we select the 4500  faintest galaxies from the
mock   group   catalogue,   using   the   same  criteria   as   described   in
Subsection~\ref{sec:samples}.   We  choose the  observational  result for  the
central  galaxies  in  S1+S2+S3  (shown   as  the  dots  with  error  bars  in
Fig.~\ref{fig:M_n} to compare  with our models.  As discussed  in the previous
section,  this result  is  representative  of the  distribution  of red  dwarf
galaxies with respect to their nearest  more massive halos.  For all Cases (I,
II and III), we adopt $f_{\rm red,  sat}=0.38$ so that the red fraction of the
satellite  galaxies  is  consistent  with  the SDSS  data  (repeated  in  Fig.
\ref{fig:mock} as a solid line).  In  Case II we set $f_{\rm red, cent}=0.07$,
so that the red fraction of  the central galaxies at large projected distance,
$\rp/R_{180}\ge  4$, is  roughly  the  same as  the  observational data.   The
corresponding  results obtained  from Case  I  and Case  II are  shown in  the
upper-left and  upper-right panels of Fig.   \ref{fig:mock}, respectively.  It
is clear  that in Case  I, in  which the only  red dwarfs are  satellites, the
fraction of  false central  galaxies in  the mock catalogue  is too  small too
match the observational data at  $\rp/R_{180}\ga 1$.  For Case II, although by
construction the  fraction of red  central galaxies matches  the observational
results at large $\rp/R_{180}\ge 4$,  the model underestimates the fraction of
red central  galaxies at  intermediate $\rp/R_{180}$.  These  results indicate
that (i) not  all red dwarf galaxies are satellites, and  (ii) the central red
dwarf  galaxies have  a different  distribution than  the total  central dwarf
population.

Now, let us  look at Case III. We have experimented  with different values for
$a$  and $b$,  and found  that  the following  set of  parameters matches  the
observational  data  reasonably   well:  $(a,b)=(0.05,  0.36)$.   The  results
obtained from the mock catalogue using  this set of model parameters are shown
in the lower-left  panel of Fig. \ref{fig:mock}.  This  shows that the central
red dwarf galaxies are correlated with massive halos on scales given by $(y-1)
\la 2$ (i.e.  $r\la 3 R_{180}$).

Finally, as  we did for the  observational sample, we can  also obtain $f_{\rm
red, cent}(r)$  for dwarf  galaxies near halos  of different masses  using the
mock sample  Case III. As an  illustration, we consider the  case with $M_n\ge
10^{12.0}\msunh$. The model  for $f_{\rm red, cent}(r)$ that  best matches the
observational  result has $(a,b)=(0.05,  0.45)$ and  is slightly  steeper than
that  obtained for  the  case without  any  restrictions on  $M_n$. The  model
prediction obtained from the mock sample  is shown in the lower-right panel of
Fig. \ref{fig:mock} along with the corresponding observational data.

In the mock catalogue considered above, the distribution of satellite galaxies
in individual halos  is assumed to be spherically symmetric  and to follow the
NFW  (Navarro,  Frenk \&  White  1997)  profile (see  Yang  et  al.  2004  for
details).  In  reality, the distribution  of satellite galaxies  in individual
halos  may not be  spherical, which  may cause  further contaminations  of the
group  memberships  selected by  the  group finder.   To  test  this, we  have
constructed  a mock  catalogue  assuming that  the  distribution of  satellite
galaxies in individual halos is triaxial,  with axis ratios given by the model
of Jing  \& Suto (2002) for  CDM halos.  We  found a slightly higher  level of
contamination in  this new mock catalogue, but  it does not change  any of our
results   significantly.     As   an   illustration,   in    each   panel   of
Fig. \ref{fig:mock}  we also  show the results  for non-spherical  halos using
symbols  connected with  dot-dashed lines.   We obtain  for Case  III slightly
weaker radial dependence with $(a,b)=(0.05,0.34)$ and $(a,b)=(0.05,0.43)$, for
the cases shown in the lower-left and lower-right panels, respectively.

\section{Discussion}
\label{sec_discussion}

We set out to understand why  about 1/4 of {\it central} dwarf galaxies, those
with $r$-band magnitudes between -14.46 and -17.05, are red when one defines a
dwarf  galaxy to be  red by  extrapolating the  division between  red sequence
galaxies and the blue cloud, as determined by Yang et al. (2008a) for brighter
galaxies, down to  dwarf magnitudes. Current models of  galaxy formation would
naively  expect such  galaxies  to be  blue  since they  would be  efficiently
accreting  gas through  cold mode  accretion  and rapidly  converting it  into
stars.

In a recent study,  Ludlow et al.  (2008; see also Lin  et al.  2003) analysed
the  properties of subhalos  in galaxy-sized  cold dark  matter halos  using a
suite  of cosmological N-body  simulations. The  subhalos in  their definition
refer to the whole population  of subhalos physically associated with the main
system, including both subhalos that are found within the virial radius of the
host halo  at the  present time, and  halos that  were once within  the virial
radius of  the main  progenitor of  the host and  have survived  as self-bound
entities until $z=0$.  They found that such populations can extend beyond {\it
three times}  the virial radius, and  contain objects on  extreme orbits, with
some  approaching the nominal  escape speed  from the  system. On  average the
subhalos identified  within the  virial radius represent  only about  {\it one
half} of all  associated subhalos, and many relatively  central halos may have
actually been ejected  in the past from a more  massive system. Since galaxies
are assumed  to form in  dark matter  halos, it is  interesting to see  if the
results we obtain here can be  understood in terms of galaxy formation in this
population of subhalos.

According to  the current theory  of galaxy formation, satellite  galaxies can
experience various environmental effects  that can quench their star formation
and make them  red (e.g., van den Bosch et al.  2008b and references therein).
Because the galaxies in ejected subhalos have also been satellite galaxies, at
least  for some period  of time,  they are  likely to  have been  subjected to
similar environmental effects, and thus  to have experienced some quenching of
their star  formation rates.  It is  thus likely that  the association  of red
dwarf  galaxies  with  massive  halos   presented  here  is  produced  by  the
association of ejected subhalos with their (former) hosts.

As shown in  Table~\ref{tab1}, about 30\% of the  dwarf galaxies are satellite
galaxies. According  to the results obtained  by Ludlow et  al.  (2008), there
should  thus  also  be a  significant  fraction  of  dwarf galaxies  that  are
physically associated with nearby more  massive halos out to about three times
the virial  radius.  If  these associated galaxies  (now outside  their hosts)
have properties similar to the satellite galaxies, we would expect an enhanced
fraction of  red dwarf  galaxies that are  distributed outside  massive halos.
This is qualitatively consistent with our findings presented above and also as
shown in Table~\ref{tab1}.  However, since the observational data are obtained
in redshift space and based on  galaxy groups that may contain interlopers and
may  be incomplete,  a detailed  comparison between  the data  and  the models
requires  the construction of  mock catalogues  that make  use of  the subhalo
population and contain all the observational selection effects.

The fact that central dwarf  galaxies have concentrations that are independent
of their distances  to the nearest massive halos  indicates that the processes
that causes  them to become  red does not  have a significant impact  on their
structure.  This  is similar to the results  obtained by van den  Bosch et al.
(2008a) who  found that the transformation mechanisms  operating on satellites
affect color more  than structure (see also Kauffmann et  al. 2004; Blanton et
al. 2005a; Ball, Loveday \& Brunner 2008; Weinmann et al.  2008).  Once again,
this  similarity between red  dwarfs that  are satellites  and those  that are
centrals suggests that both populations  may have experienced similar kinds of
environmental effects.

However, as shown  in Table~\ref{tab1}, in the S1+S2+S3  sample less than 42\%
of the  red dwarf central galaxies  have $\rp/R_{180}\le 3$.   In other words,
more than  58\% of the red  dwarf central galaxies  are not close enough  to a
larger halo so that they could  have been preprocessed there, becoming red and
then subsequently being ejected.  The origin of this population of central red
dwarf galaxies,  which is almost 10\%  of the combined  S1+S2+S3 dwarf sample,
still  remains a  mystery within  the standard  paradigm of  galaxy formation.
Furthermore,  if we had  used stellar  mass instead  of $r$-band  magnitude to
define our dwarf  sample, the percentage of galaxies  in this population would
likely increase.  This population of isolated red dwarfs are not merely a dust
reddened star forming  population seen near edge-on because  their axis ratios
are  consistent with  a randomly  oriented population.  Although as  Croton \&
Farrar  (2008)  probed  the origin  of  the  red  dwarfs  in voids  using  the
semi-analytical  models, they only  found $\sim  0.4\%$ of  the dwarfs  in the
total population  are red  centrals in  voids. While here  we find  that $\sim
10\%$ dwarfs  are red centrals  without close neighbours  with $\rp/R_{180}\le
3$.

An outstanding problem  for all galaxy formation models  concerns the low mass
slope of  the galaxy mass  function.  CDM models  in general predict  too many
low-mass dark matter haloes compared to  the number of low mass galaxies.  The
mass function of dark matter haloes,  $n(M)$, scales with halo mass roughly as
$n(M)\propto M^{-2}$ at the low-mass end.  This is in strong contrast with the
observed  luminosity function  of galaxies,  $\Phi  (L)$, which  has a  rather
shallow shape  at the faint end,  with $\Phi(L) \propto  L^{-1}$. To reconcile
this difference  one usually  invokes some form  of feedback within  these low
mass halos.  If the feedback mechanism were to prevent gas from entering these
halos  at  late times,  such  galaxies would  appear  red.   For example,  the
preheating mechanism of Mo et al. (2005), where gas is preheated by gas shocks
within the  forming large scale structures  in which the low  mass dark matter
halos themselves are forming, has this feature.  Therefore, this population of
isolated, red, central dwarf galaxies  could represent the tail of the process
that prevents  the vast majority  of low mass  dark matter halos  from forming
galaxies  and  their further  study  could shed  new  light  on the  mechanism
responsible.

%%%%%%%%%%%%%%%%%
% Ackowledgements
%%%%%%%%%%%%%%%%%

\acknowledgements We thank the referee Darren Croton for helpful comments that
improved the presentation of this  paper. YW acknowledges the support of China
Postdoctoral  Science Foundation.  This  work  is supported  by  the {\it  One
Hundred  Talents}  project,  Shanghai  Pujiang Program  (No.  07pj14102),  973
Program (No.   2007CB815402), the CAS Knowledge Innovation  Program (Grant No.
KJCX2-YW-T05) and  grants from NSFC (Nos.  10533030,  10673023, 10821302). HJM
would like to acknowledge the support of NSF AST-0607535, NASA AISR-126270 and
NSF IIS-0611948.  NSK and D.H.M. would like to acknowledge the support of NASA
LTSA NAG5-13102.

%%%%%%%%%%%%%%%%%
% Bibliography
%%%%%%%%%%%%%%%%%

\clearpage


\begin{thebibliography}{}


\bibitem{Ade06} Adelman-McCarthy J.K., et al., 2006, \apjs, 162, 38

\bibitem{Bal04} Baldry I.K., Glazebrook K., Brinkmann J., Ivezic Z., Lupton
  R.H., Nichol R.C., Szalay A.S., 2004, \apj, 600, 681

\bibitem{ball} Ball N.M., Loveday J., Brunner R.J., 2008, \mnras, 383, 907

\bibitem{BM00} Balogh M.L., Morris S.L., 2000, \mnras , 318, 703

\bibitem{Bal00} Balogh M.L., Navarro J.F., Morris S.L., 2000, \apj , 540, 113

\bibitem{Barr84} Barrow, J. D., Bhavsar, S. P., Sonoda, D. H., 1984, \mnras,
  210, 19

\bibitem{Bell03} Bell E.F., McIntosh D.H., Katz N., Weinberg M.D., 2003, ApJS,149,
289

\bibitem{Benson 2005} Benson, A. J., \ 2005, \mnras, 358, 551

\bibitem{Bla05} Blanton M.R., Eisenstein D.J., Hogg D.W., Schlegel D.J.,
  Brinkmann J., 2005a, \apj, 629, 143

\bibitem{NYUVAGC} Blanton M.R. \etal, 2005b, \aj, 129, 2562

\bibitem{Brown08} Brown M.J.I., \etal, 2008, \apj, 682, 937

\bibitem{Cac08} Cacciato M., van den Bosch F. C., More S., Li R., Mo H. J.,
  Yang X., \ 2008, preprint, arXiv:0807.4932

\bibitem{Coop07} Cooper M.C., et al., 2007, \mnras, 376, 1445

\bibitem{Cro08} Croton D. J.; Farrar G. R., 2008, \mnras, 386, 2285

\bibitem{Diemand et al 07a} Diemand, J., Kuhlen, M. Madau, P., \ 2007, \apj,
  657, 262

\bibitem{Diemand et al 07b} Diemand, J., Kuhlen, M. Madau, P., \ 2007, \apj,
  667, 859

\bibitem{Dressler et al 07b} Dressler A., et al., 1997, \apj, 490, 577

\bibitem{Gao et al 2004} Gao, L., White, S.D.M., Jenkins, A., Stoehr, F.,
  Springel, S., \ 2004, \mnras, 355, 819G

\bibitem{Ghigna et al 1998} Ghigna, S., Moore, B., Governato, F., Lake,
  G. Quinn, T., Stadel, J., \ 1998, \mnras 300, 146

\bibitem{Ghigna et al 1999}
Ghigna, S., Moore, B., Governato, F., Lake, G. Quinn, T., Stadel, J.,
\ 1999, aspc 176, 140

\bibitem{Gio08} Giocoli C., Tormen G., van den Bosch F.C., 2008, \mnras, 386,
  2135

\bibitem{Gun72} Gunn J.E., Gott J.R., 1972, \apj , 176, 1

\bibitem{Guo09} Guo Y., McIntosh D. H., Mo H. J., Katz Neal, van den
Bosch F. C., Weinberg M., Weinmann S. M., Pasqual A., Yang X., 2009,
arXiv0901.1150

\bibitem{Jing} Jing Y.P., Suto Y., 2002, \apj, 574, 538 (JS02)

\bibitem{Kau93} Kauffmann G., White S.D.M, Guiderdoni B., 1993, \mnras, 264,
  201

\bibitem{Kau03}
Kauffmann G., Heckman T.M., White S.D.M, Charlot S., Tremonti, C., Peng E.W.,
Seibert M., Brinkmann J., Nichol R.C., SubbaRoa M., York D., 2003, \mnras, 341,
54

\bibitem{kauffmann2} Kauffmann G., White S.D.M, Heckman T.M., M\'{e}nard B.,
  Brinchmann J., Tremonti C., Brinkmann J., 2004, \mnras, 353, 713

\bibitem{Keres05} Keres, D., Katz, N., Weinberg, D.H., \& Dave\'e, R., 2005,
  \mnras, 363,2

\bibitem{Keres08}
Keres, D., Katz, N., Fardal, M.,  Dave\'e, R., \& Weinberg, D.H., 2008,
\mnras, submitted (arXiv:0809.1430)

\bibitem{Lar80} Larson R.B., Tinsley B.M., Caldwell C.N., 1980, \apj , 237,
  692

\bibitem{Li06} Li C., Kauffmann G., Jing Y.P., White S.D.M., B\"orner G.,
  Cheng F.Z., 2006, \mnras , 368, 21

\bibitem{Li07} Li C., Jing Y.P., Kauffmann G., B\"orner G, Kang X., Wang L.,
  2007, \mnras , 376, 984

\bibitem{Lin03} Lin W.P., Jing Y.P., Lin L., 2003, \mnras, 344, 1327

\bibitem{Ludlow et al. 2008} Ludlow A.D., Navarro J.F., Springel V., Jenkins
  A., Frenk C.S., Helmi A., 2008, preprint, arXiv:0801.1127

\bibitem{Mo92}
Mo H. J., Jing Y.P., B\"orner G., 1992, \apj, 392, 452

\bibitem{Mo05}
Mo H. J., Yang X., van den Bosch F.C., Katz N., 2005, \mnras, 363,
1155

\bibitem{Moore et al. 1996}
Moore, B., Katz, N., Lake, G., Dressler, A. \& Oemler, A., 1996,
Nature, 379, 613

\bibitem{Moore et al. 1999} Moore, B., Ghigna, S., Governato, F., Lake, G.,
  Quinn, T., Stadel, J., Tozzi, P., \ 1999, \apj, 5 24, L19

\bibitem{NFW} Navarro J.F., Frenk C.S., White S.D.M., 1997, \apj, 490, 493

\bibitem{PH} Park C., Hwang H.S., 2008, preprint, arXiv:0812.2088

\bibitem{Qui00} Quilis V., Moore B., Bower R., 2000, Science, 288, 1617

\bibitem{Spe07} Spergel D. N., et al., 2007, ApJS, 170, 377

\bibitem{Str01} Strateva I., et al., 2001, AJ, 122, 1861

\bibitem{B05} van den Bosch F.C., Tormen G., Giocoli C., 2005, \mnras , 359,
  1029

\bibitem{B08a} van den Bosch F. C., Pasquali, A., Yang X., Mo H. J., Weinmann
  S.M., McIntosh D. H.; Aquino D., 2008a, preprint, arXiv:0805.0002

\bibitem{B08b} van den Bosch F.C., Aquino D., Yang X, Mo H.J., Pasquali A.,
  McIntosh D.H., Weinmann S.M., Kang X., 2008b, \mnras, 387, 79

\bibitem{Wang08}Wang H. Y., Mo H. J., Jing Y. P., 2008, preprint,
  arXiv:0811.3558

\bibitem{Wei08} Weinmann S. M., Kauffmann G., van den Bosch F. C., Pasquali
  A., McIntosh D. H., Mo H., Yang X., Guo Y., 2008, preprint, arXiv:0809.2283

\bibitem{Y04} Yang X., Mo H.J., Jing Y.P., van den Bosch F.C., Chu Y.Q., 2004,
  \mnras , 350, 1153

\bibitem{Y05} Yang X., Mo H.J., van den Bosch F.C., Jing Y.P., 2005, \mnras,
  356, 1293

\bibitem{Y07} Yang X., Mo H.J., van den Bosch F.C., Pasquali A., Li C., Barden
  M., 2007, \apj, 671, 153 (Paper I)

\bibitem{YMB08a} Yang X., Mo H.J., van den Bosch F.C., 2008a, \apj, 676, 248

\bibitem{YMB08b} Yang X., Mo H.J., van den Bosch F.C., 2008b, preprint,
  arXiv:0808.0539

\bibitem{YMB08c} Yang X., Mo H.J., van den Bosch F.C., 2008c, preprint,
  arXiv:0808.2526

\end{thebibliography}
\end{document}